\documentclass[aps,twocolumn]{revtex4-1}
\usepackage{graphicx}
\usepackage{color}
\usepackage{amssymb}
\usepackage{bm}

\newcommand{\sgn}{{\rm sgn}}

\begin{document}

\title{Fermi Surface Resonance and Quantum Criticality in Strongly Interacting Fermi Gases} 

\date{\today}
\author{Dmitry Miserev,$^{1\ast}$ Jelena Klinovaja,$^{1}$ and Daniel Loss$^{1}$}
\affiliation{$^{1}$Department of Physics, University of Basel, \\
	Klingelbergstrasse 82, CH-4056 Basel, Switzerland\\
}

\begin{abstract}
Fermions in a Fermi gas obey the Pauli exclusion principle restricting any two fermions from occupying the same quantum state. Strong interactions between fermions can completely change the properties of the Fermi gas. In our theoretical study we find a new exotic quantum phase in strongly interacting Fermi gases subject to a certain condition imposed on the Fermi surfaces that we call the Fermi surface resonance.	The new phase is quantum critical in time and space and can be identified by the power-law  dependence of the spectral density in frequency and momentum. The linear response functions are singular in the static limit and at the Kohn anomalies. We analyze the quantum critical state at finite temperatures $T$ and finite size $L$ of the Fermi gas and provide a qualitative $L$-$T$ phase diagram.
The new quantum critical phase can be experimentally found in typical semiconductor heterostructures.
\end{abstract}

\maketitle

\section{Introduction}

Physical properties of  Fermi gases in a large variety of different materials have been extensively studied over the past century \cite{Ashcroft}. Properties of the non-interacting Fermi gas are entirely determined by single-particle physics and the Fermi statistics \cite{Fermi}. Fermions in condensed matter physics are represented by electrons or holes that interact via the Coulomb force.
The Coulomb interaction between fermions can significantly change the properties of a Fermi gas.
For example, a one-dimensional Fermi gas forms the strongly correlated Tomonaga-Luttinger liquid at arbitrarily weak interactions \cite{Tomonaga,Luttinger,Mattis}.
In higher spatial dimensions, however, weak interactions do not spoil the properties of Fermi gases
but only slightly change the non-interacting characteristics. The gas of such weakly interacting fermions can be modeled by the gas of ``dressed'' non-interacting Landau quasiparticles. The gas of Landau quasiparticles is known as the Landau Fermi liquid (LFL) \cite{Landau}.

If the interaction is strong, the LFL breaks down and the ground state of the strongly interacting fermion system can dramatically change.
The interaction strength is measured by the dimensionless interaction parameter $r_s$:
\begin{eqnarray}
\displaystyle r_s = \frac{v_C}{E_F} \sim \frac{m e^2}{\epsilon n^{\frac{1}{D}}} ,
\label{rs}
\end{eqnarray}
where $v_C$ is the Coulomb interaction (on average), $E_F$ the Fermi energy, $n$ the fermion density, $m$ the effective mass, $e$ the elementary charge, $\epsilon$  the dielectric constant, and $D$ the spatial dimension.
Thus, in order to drive the system into the strongly interacting regime $r_s \gg 1$, we generally need a large effective mass $m$, small dielectric constant $\epsilon$ and small density $n$.
For example, the strongly interacting electron gas in near magic angle twisted bilayer graphene exhibits exotic magnetism \cite{Kotov}, charge density order \cite{Jiang}, and unconventional superconductivity \cite{Cao}, because $r_s \gg 1$ due to the low electron density and large effective mass.
The hole-doped semiconductors such as GaAs, InAs, InSb \cite{Ando}, or Ge \cite{Scappucci}
are also good candidates because of the large effective hole mass. 
The hole density $n$ can be tuned to sufficiently small values by the electrostatic gates.
Taking a two-dimensional semiconductor  \cite{Ando} with $\epsilon = 10$, $m = 0.2 m_0$, where $m_0$ is the bare electron mass, and $n = 10^{11}$ cm$^{-2}$,
we get $r_s \sim 10 \gg 1$, which corresponds to the strongly interacting regime.

The Coulomb interaction between charged fermions can be divided into two physically different parts.
The first one is the classical electrostatic interaction with other electric charges via the charge density.
In quantum physics there is one more type of  interaction which comes from the quantum indistinguishability of two interacting fermions of the same type. This is the exchange interaction \cite{Heisenberg}. The exchange interaction can mix quasiparticles from different Fermi surfaces. In our study we show that, under certain conditions on the Fermi surfaces, the exchange interaction mixes the fermions into a new exotic phase.
In this new phase the fermions form a strongly interacting quantum liquid and the LFL quasiparticle picture breaks down.

In the absence of quasiparticles there is no simple visual picture to characterize quantum processes.
In order to describe quantum liquids with no quasiparticles, a quantum field description is required \cite{Fradkin}.
Excitations or quanta of the fermion fields in the LFL are long lived and they represent the Landau quasiparticles.
The Heisenberg uncertainty principle obliges all physical fields to fluctuate. For example, quantum fluctuations in the LFL result in the finite lifetime of the Landau quasiparticles \cite{Pomeranchuk}. However, quantum fluctuations in strongly interacting quantum liquids completely destroy quasiparticles \cite{Sachdev}. This means that all field excitations are strongly damped by the quantum fluctuations and cannot be considered as sharply defined long lived resonances. The single-particle methods in such quantum liquids are inadequate and, instead, the fermion correlations must be considered.

In this work we calculate the fermion Green function. The Green function is connected to some observables, e.g. to the spectral function and to linear response functions such as conductivity and spin susceptibility.
The spectral function in the new phase contains no quasiparticle poles and instead shows universal power-law scaling with frequency and momentum. The linear response functions are singular in the static limit and at the Kohn anomalies. Strongly interacting quantum liquids with these properties are called quantum critical \cite{Coleman}.

Quantum criticality in itinerant electron gases has been considered previously in application to cuprates  \cite{Daou,Keimer,Gegenwart}. Corresponding theoretical models are either based on the coupling of fermions 
to a bosonic critical order parameter \cite{chubukovye,chubukov,vojta}, or on proximity to the Mott transition \cite{senthilmott},
or on the Sachdev-Ye-Kitaev (SYK) \cite{Ye,SachdevX}
scenario of  quantum criticality \cite{Parcollet,Chowdhury,Davison}.
The SYK model describes flat band fermions with long range all-to-all interactions whose matrix elements are randomly distributed. In our model,  we do not couple fermions to a critical order parameter, neither do we consider the Mott transition nor require random or even long-range interaction. The quantum criticality in our model emerges due to a resonant many-body exchange interaction between electrons that belong to different Fermi surfaces. This resonance, referred to as Fermi surface resonance (FSR), is the crucial feature of our model with far reaching consequences, both theoretically and experimentally,
as explained in great detail in the following sections.

 This work is organized as follows.
In Sec. II we describe the FSR and 
provide an example of a realistic physical system which can be tuned to the FSR.
In Sec. III we construct the effective Hamiltonian which is sensitive to the FSR.
In Sec. IV we calculate the electron self-energy
corresponding to the effective Hamiltonian.
In Sec. V we investigate the strong coupling limit 
which is characterized by the emergent temporal and spatial conformal symmetry.
The temporal part of the Green function is shown to obey the SYK equation, while
the spatial part obeys a generalized version of it.
In Sec. VI we fix the interaction strength 
and consider the crossover between the LFL 
and the quantum critical state on the $L$-$T$ diagram,
where $T$ is the temperature and $L$  the sample size. The linear response functions in the quantum critical state are calculated in Sec. VII. We find a temperature behavior of the resistance similar to the one found in strange metals. Conclusions are given in Sec. VIII. Some technical details are deferred to two appendices.

\section{Fermi surface resonance}

\begin{figure}[t]
	\includegraphics[width=\columnwidth]{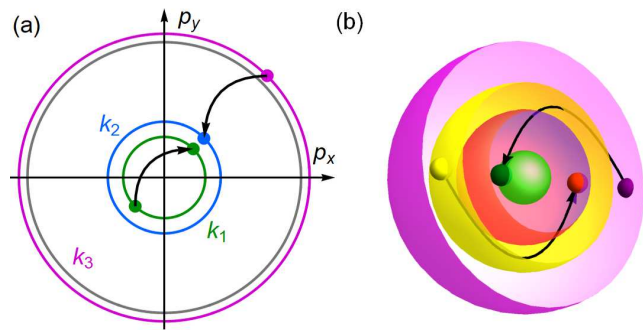}
	\caption{ 
		Examples of the FSR.
		Some of the elementary resonant processes (not all of them) are shown by the arrows.
		(a)
		The simplest example of FSR 
		can be realized in semiconductor heterostructures with two filled subbands, see Eq.~(\ref{example}).
		The spin degeneracy of the subbands is lifted by the applied magnetic field.
		The resonant Fermi surfaces are indicated by color.
		The gray Fermi surface is off resonance.
		(b)
		3D example of the FSR. 
		Only the resonant Fermi surfaces are shown.
	}
\label{fig:ex}
\end{figure}

In our study we consider a Fermi gas with multiple non-degenerate Fermi surfaces.
This can be experimentally realized in semiconductor heterostructures \cite{Dresselhaus}.
Semiconductor heterostructures consist of  thin semiconductor layers.
The contact potential between the layers confines electrons or holes within one layer.
This leads to  quantized energy subbands corresponding to different confined modes.
Filling multiple subbands results in multiple Fermi surfaces, see Fig.~1.
Generally, the Fermi surfaces are degenerate due to spin.
The spin degeneracy can be lifted by an applied in-plane magnetic field or by spin-orbit interaction \cite{Manchon}.
The spin-orbit interaction can be precisely tuned by electric gates \cite{Zumbuhl}.
We assume that $2 N$, $N \ge 2$, of the non-degenerate Fermi surfaces can be tuned close to the FSR:
\begin{eqnarray}
\mathcal{K} = 0 ,
\label{reducedconserv}
\end{eqnarray}
where $\mathcal{K}$ is a momentum mismatch between the corresponding Fermi momenta $k_a > 0$ of the involved Fermi surfaces (we assume $\hbar=1$ throughout), 
\begin{eqnarray}
\mathcal{K} = k_1 + \dots + k_{N + \sigma} - k_{N + \sigma + 1} - \dots - k_{2 N} .
\label{Delta}
\end{eqnarray}
Here, $\sigma \in \{0, 1,  \dots, N - 1\}$.
Not all Fermi surfaces are required to take part in the resonance, e.g. see Fig.~\ref{fig:ex}(a).
The Fermi surfaces are assumed to be spherically symmetric.
This is often the case in semiconductor heterostructures because the electron and hole dispersions at small densities are nearly isotropic \cite{Kriisa}.

Equation~(\ref{reducedconserv}) can be thought of as a radial nesting of the Fermi surfaces.
Generally, nesting implies a one-dimensional character of the scattering between the nested parts of the Fermi surface.
This leads to strong enhancement of such scattering which can trigger an instability.
Celebrated examples of  instabilities driven by nesting are the charge and spin density orders \cite{Whangbo,Murayama}.
The radial nesting is also known to result in strongly interacting electron states such as fractional topological insulators with a gap~\cite{Trifunovic,Volpez}.
In our study we show that the radial nesting of the Fermi surfaces given by Eq.~(\ref{reducedconserv}) leads to a gapless quantum critical state.

In the general formulation of the problem we require all $2 N$ non-degenerate Fermi surfaces participating in the resonant $N \to N$ scattering [see Eqs.~(\ref{reducedconserv}), (\ref{Delta})] to be different.
However, we can soften this and only require
that the $N$ initial states belong to different Fermi surfaces,
and similarly for the $N$ final states; some of the initial states might then have the same Fermi surface index as some of the final states.
In the example shown in Fig.~\ref{fig:ex}(a)
the FSR corresponds to the following condition:
\begin{eqnarray}
2 k_1 + k_2 - k_3 = 0 ,
\label{example}
\end{eqnarray} 
where $k_a$, again, are the Fermi momenta of corresponding Fermi surfaces.
The gray Fermi surface in Fig.~\ref{fig:ex}(a) does not participate in the resonant scattering.
This example corresponds to the soft formulation because the green Fermi surface contains initial and final states.
The integer $N$ in the soft formulation corresponds to the $N$-particle resonant scattering amplitude, e.g. $N = 2$ in Fig.~\ref{fig:ex}(a).

In order to study the new quantum critical phase experimentally, one has to satisfy the single resonant condition given by Eq.~(\ref{reducedconserv}).
In addition, one has to ensure that the Fermi gas is strongly interacting, i.e. $r_s \gg 1$.
We argue that  semiconductor heterostructures are most promising candidates for the experimental search for such new phases.
The simplest example of a semiconductor heterostructure that can host a new exotic quantum critical phase is shown in Fig.~\ref{fig:ex}(a), which represents
the Fermi surfaces of a two-dimensional electron gas with two occupied subbands.
Even though the occupation of multiple subbands is experimentally achievable \cite{Hamilton,Zheng}, 
the experimental research of materials with multiple Fermi surfaces is still very limited which partially explains why this new phase has never been detected before.
Each subband in Fig.~\ref{fig:ex}(a) is split by an external magnetic field.
Changing the electron density and fine tuning by the magnetic Zeeman splitting we can set the system to the FSR given by Eq.~(\ref{example}).
The given example corresponds to the $N = 2$ particle resonant scattering amplitude within the soft formulation of the FSR
because the green Fermi surface contains both initial and final states.
The FSR results in the strong mixing of three colored bands in Fig.~\ref{fig:ex}(a) which destroys quasiparticles in the vicinity of the colored Fermi surfaces.
The quasiparticles in the vicinity of the off-resonant gray Fermi surface survive.
The new phase in this example has separate Fermi-liquid and non-Fermi-liquid components.
The latter is established at the resonance given by Eq.~(\ref{example}) and can be experimentally identified from the power-law frequency and momentum dependence of the spectral function
and the singular linear response functions in the static limit and at the Kohn anomalies (see below).

In what follows we consider the general case of $D > 1$ spatial dimensions.
A $D=3$ example of the strong version of the FSR with $2 N = 4$ different Fermi surfaces 
is shown in Fig.~\ref{fig:ex}(b).

\section{Effective Hamiltonian}

Now we proceed to the general case of the FSR.
Here we assume that all $2 N$ fields participating in the resonance are different, see Eq.~(\ref{reducedconserv}).
All the results that we obtain in this paper also apply to the soft version of the FSR where some initial states might have the same Fermi surface index as some final states, see Fig.~\ref{fig:ex}(a).
The FSR results in a dramatic change of the ground state because it favors resonant many-body exchange scattering.
The FSR is applied to $2 N$ different non-degenerate Fermi surfaces, so we consider an $N \to N$ scattering amplitude which is multilinear with respect to each fermion field.
As the FSR condition says nothing about initial and final states, we have to sum over all possible choices of $N$ initial and $N$ final states out of overall $2 N$ fields corresponding to the $2 N$ Fermi surfaces, 
yielding the following effective Hamiltonian:
\begin{eqnarray}
\!\!\!\!\! V \! (R) = \! \sum\limits_{\{j\}} \lambda_j \Psi_{j_1}^\dagger \! (R) \dots \Psi_{j_N}^\dagger \! (R) \Psi_{j_{N + 1}} \! (R)  \dots \Psi_{j_{2 N}} \! (R) ,
\label{VN}
\end{eqnarray}
where $R = (t, {\bf r})$, ${\bf r}$ is a $D$-dimensional position vector, $t$ the time, $j$  a permutation of indices $\{1, \dots, 2 N\}$, $\Psi_a (R)$  the fermion field operator corresponding to the $a^{\mathrm{th}}$ Fermi surface, and $\lambda_j$ are the coupling constants.
We sum over all non-equivalent permutations corresponding to $C^N_{2 N} = (2 N)! / (N!)^2$ different choices of initial and final states. 
For conjugate terms the corresponding $\lambda_j$ is a complex conjugate in order to ensure hermiticity of $V (R)$.
The effective Hamiltonian $V (R)$ is of exchange form as it mixes together all $2 N$ fermion fields corresponding to the $2 N$ Fermi surfaces participating in the resonance.
Similar in spirit is the effective Hamiltonian approach  widely used in condensed matter physics, in particular, in the weakly coupled wire approach where $N$-electron effective inter-wire interactions are constructed \cite{Teo,Aseev,Klino}.

In case of the soft formulation, the effective Hamiltonian has the same form as Eq.~(\ref{VN}). 
The only difference is that the terms containing the square of field operators vanish due to the Fermi statistics.
This is consistent with our requirement that all $N$ initial states as well as all $N$ final states   belong to $N$ different Fermi surfaces.

The effective Hamiltonian, see Eq.~(\ref{VN}), can be constructed using  perturbation theory with respect to the two-particle Coulomb interaction, $v_C$.
The first contribution to $V (R)$ comes from the tree diagrams in the $(N-1)^{\mathrm{th}}$ order in $v_C$:
\begin{equation}
V(R) \propto \Lambda = \frac{v_C^{N - 1}}{E_F^{N - 2}} = E_F r_s^{N - 1} ,
\label{Lambda}
\end{equation} 
where $\Lambda$ is the characteristic energy scale of $V (R)$, $E_F$  the Fermi energy, and $r_s$  the interaction parameter given by Eq.~(\ref{rs}).
The power of $v_C$ corresponds to the order of perturbation theory, the power of $1/E_F$ corresponds to the number of fermion propagators in the tree diagrams.
Notice that the strongly interacting regime $r_s \gg 1$ also corresponds to $\Lambda \gg E_F$.
Each of the tree diagrams can be envisioned as a sequence of $N - 1$ Coulomb exchange scattering events.
As all $2 N$ fermions are different, the momentum transfers during the exchange are all in order of the average Fermi momentum $k_F$.
We are interested in the long range correlations at relative distance $r \gg 1 / k_F$.
For such long range correlations the $N \to N$ scattering that occurs on the scale of the Fermi wavelength $\sim 1 / k_F$ is effectively local, justifying the locality of $V(R)$.
All the matrix elements that appear in the tree diagrams for $V (R)$ are hidden in the coupling constants $\lambda_j$.
Higher order diagrams for $V(R)$ only renormalize the coupling constants $\lambda_j$.
Due to symmetries of specific Hamiltonians some of the coupling constants $\lambda_j$ might be equal to zero.
However, this fact is not important for the further analysis if there are at least some non-zero $\lambda_j$.

We argue that $V(R)$ is the most important scattering amplitude close to the FSR, see Eq.~(\ref{reducedconserv}).
All other terms in the many-body scattering amplitudes are either insensitive to the FSR or contain rapidly oscillating terms on the scale of Fermi wavelength.
Here we work within the assumption that at arbitrary filling electrons or holes in semiconductors form the LFL.
This means that the interaction which is not sensitive to the FSR cannot significantly change the physics.
Interactions that contain oscillating terms can be averaged to zero on large scales $r \gg 1 / k_F$.
This allows one to include such interactions as irrelevant corrections renormalizing the LFL parameters.

The effective Hamiltonian $V(R)$ is very sensitive to the FSR condition given by Eq.~(\ref{reducedconserv}).
Further we show that 
exactly at the FSR it results in the emergent long-range order 
and the destruction of the quasiparticle picture 
in the infrared limit.

\section{Dimensional reduction}

In this section we calculate the fermion Green function dressed by the effective interaction $V(R)$, see Eq.~(\ref{VN}).
Here we show that a strong interaction $V(R)$ completely destroys quasiparticles close to the Fermi surfaces.
The notion of Fermi surfaces is still important though because they define the sector of quantum states that are  most affected by the interaction $V(R)$.
Such Fermi surfaces without quasiparticles are known as critical Fermi surfaces \cite{Chowdhury}.

Far away from the Fermi surfaces 
the resonant scattering is destroyed and 
thus we expect the LFL in the ultraviolet limit
even when the FSR condition Eq.~(\ref{reducedconserv})
is satisfied.
Thus, for large $\omega_a \sim E_F$ and $\delta p_a = p_a - k_a \sim k_a$ 
the Green function restores the quasiparticle poles:
\begin{eqnarray}
G^{(0)}_a (\omega_a, \delta p_a) = \frac{Z_a}{\omega_a - v_a \delta p_a} ,
\label{G0}
\end{eqnarray}
where $E_F$ is the Fermi energy,
the index $a \in \{1, \dots, 2 N\}$ enumerates the Fermi surfaces,
$\omega_a$ and $p_a$ are the frequency 
and the momentum, 
$k_a$ is the Fermi momentum of the $a^{\rm th}$ Fermi surface,
$v_a$ is the Fermi velocity which is renormalized by irrelevant interactions, and
$1 > Z_a > 0$ is the quasiparticle residue away from the Fermi surface.
We allow the frequency $\omega_a$ to be complex.
For example, 
imaginary frequencies correspond to the Matsubara formalism,
while $\omega_a \to \omega_a + i0$ yields the 
retarded Green function.
In this section we calculate the Matsubara Green function at zero temperature.
Other Green functions can be obtained 
via the analytical continuation
through the spectral representation \cite{lehmann}:
\begin{eqnarray}
G_a (\omega_a, \delta p_a) = \int\limits_{-\infty}^\infty  \frac{dz}{\pi}\frac{ \mathcal{A}_a (z, \delta p_a)}{\omega_a - z} ,
\label{specrep}
\end{eqnarray}
where $\mathcal{A}_a (\omega_a, \delta p_a) = - \mbox{\rm{Im}}\left[G_a (\omega_a + i0, \delta p_a)\right] > 0$ is the positively defined spectral function, 
and $\mbox{\rm{Im}}$ stands for the imaginary part.

Close to the FSR, see Eqs.~(\ref{reducedconserv}) and (\ref{Delta}),
the resonant many-body exchange scattering described by the effective Hamiltonian $V(R)$, see Eq.~(\ref{VN}),
becomes important in the infrared limit.
We include it via the self-energy $\Sigma_a (i \omega_a, \delta p_a)$:
\begin{eqnarray}
G_a (i \omega_a, \delta p_a) = \frac{1}{G^{(0)}_a (i \omega_a, \delta p_a)^{-1} - \Sigma_a (i \omega_a, \delta p_a)} ,
\label{G}
\end{eqnarray}
where $G^{(0)}_a (i \omega_a, \delta p_a)$ is given by Eq.~(\ref{G0}) and contains the contributions from the irrelevant interactions,
$\omega_a$ is the fermionic Matsubara frequency.
We assume here that there is no Pomeranchuk instability \cite{pomeranchukinsta} even in the strongly interacting regime,
so the spherical symmetry of the Fermi surfaces is exact.
In this case the fermion Green function $G_a (\tau, \bm r)$ in the imaginary time-coordinate representation depends 
only on the absolute value  $r=|\bm r|$.
This allows for the effective one-dimensional representation of the Green function:
\begin{eqnarray}
&& G_a(\tau, r) = \sum\limits_{\bm p_a} e^{i \bm p_a \cdot \bm r} G_a(\tau, \delta p_a) \nonumber \\
&& = \int\limits_{-k_a}^\infty \frac{d \delta p_a}{2 \pi} p_a^{D - 1} J(r p_a) G_a(\tau, \delta p_a) , \label{Gint} \\
&& \sum\limits_{\bm p_a} \equiv \int \frac{d \bm p_a}{(2 \pi)^D} = \int \frac{d \Omega_D}{(2 \pi)^{D - 1}} \frac{d \delta p_a}{2 \pi} p_a^{D - 1} , \label{pint} \\
&& J (z) = \int \frac{d \Omega_D \, e^{i z \cos \theta}}{(2 \pi)^{D - 1}} \to \frac{2 \cos \left(|z| - \frac{\pi}{2} \frac{D - 1}{2}\right)}{\left(2 \pi |z|\right)^{\frac{D - 1}{2}}} ,
\label{J}
\end{eqnarray}
where $\delta p_a = p_a - k_a$, 
$k_a$ is the Fermi momentum of the $a^{\rm th}$ Fermi surface,
$d \Omega_D$ is the volume element of the $D$-dimensional solid angle.
In Eq.~(\ref{J}) we also provide the asymptotic behavior of $J(z)$ for large argument $|z|\gg 1$.
We use this to derive the long distance form of the Green function:
\begin{eqnarray}
&& G_a(\tau, r) \to \left(\frac{k_a}{2 \pi r}\right)^{\frac{D - 1}{2}}  \nonumber \\
&&\times \left[i^{-\frac{D - 1}{2}}  e^{i k_a r} \mathcal{G}_a(\tau, r) + i^{\frac{D - 1}{2}}  e^{-i k_a r} \mathcal{G}_a(\tau, -r) \right], \label{Gred}\\
&& \mathcal{G}_a(\tau, x) = \int\limits_{-\infty}^\infty \frac{d \delta p_a}{2 \pi} e^{i x \delta p_a} G_a(\tau, \delta p_a) , \label{gonedim}
\end{eqnarray}
where $r \gg 1/k_a$, 
and the integral over $\delta p_a$ is extended to the interval $(-\infty, \infty)$ with negligible error.
Here, $\mathcal{G}_a(\tau, x)$ represents a one-dimensional Fourier transform, $x \in (-\infty, \infty)$.
Thus, one can consider Eqs.~(\ref{Gred})--(\ref{gonedim})
as the dimensional reduction from $D$ spatial dimensions to a single spatial dimension with coordinate $x$
which is conjugate to the momentum $\delta p_a$.

\begin{figure}[t]
	\includegraphics[width=\columnwidth]{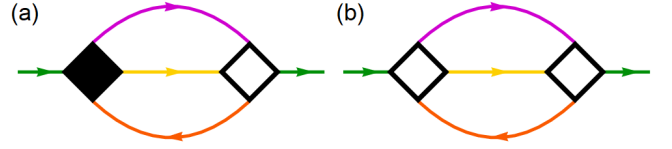}
	\caption{ 
		Electron self-energy.
		(a) Feynman diagram for the exact self-energy for the case $N = 2$.
		Solid lines correspond to the exact Green functions, see Eq.~(\ref{G}).
		The Fermi surface index $a \in \{1,2,3,4\}$ is indicated by color and corresponds to the example given in Fig.~\ref{fig:ex}(b).
		The black (white) square corresponds to the exact (bare) interaction vertex.
		There are two more contributions to the self-energy that differ by the arrow directions.
		The Feynman diagrams for general $N$ can be drawn similarly.
		(b) The SCBA. 
		The interaction vertex correction is neglected which is shown by two bare vertices (white squares).
	}
	\label{fig:se}
\end{figure}

In order to calculate the effectively one-dimensional Green function $\mathcal{G}_a(\tau, x)$, see Eq.~(\ref{gonedim}),
we have to find the electron self-energy
due to the interaction Hamiltonian $V(R)$, see Eq.~(\ref{VN}).
The Feynman diagram for the exact self-energy for $N = 2$ is presented in Fig.~\ref{fig:se}(a)
and corresponds to the example shown in Fig.~\ref{fig:ex}(b).
Feynman diagrams for general $N$ can be drawn in a similar fashion.
The problem here is the renormalization of the interaction vertex, see full black square in Fig.~\ref{fig:se}(a).
In this section we omit the interaction vertex renormalization and instead consider the simpler diagram in Fig.~\ref{fig:se}(b).
Such an approximation is called the self-consistent Born approximation (SCBA).
The diagrams of the form in Fig.~\ref{fig:se}(b) are also known as ``melon'' diagrams that appear in various matrix and tensor field theories \cite{Gurau}.

We calculate the SCBA self-energy $\Sigma_1 (\tau, \delta p_1)$, see Fig.~\ref{fig:se}(b),
in the imaginary time-coordinate representation:
\begin{eqnarray}
&& \Sigma_1 (\tau, r) = (-1)^{N - 1} \sum\limits_{\{\eta\}} |\lambda_\eta|^2 \prod\limits_{a = 2}^{2 N} G_{a} (\eta_a \tau, r) , \label{sigmageneral}
\end{eqnarray}
where $\lambda_\eta$ are the bare interaction coupling constants, see Eq.~(\ref{VN}).
We consider the self-energy corresponding to the electrons near the 1$^{\mathrm{st}}$ Fermi surface, 
the result is similar for other Fermi surface indices.
Each $\eta = (\eta_2, \dots, \eta_{2 N})$, $\eta_b = \pm 1$, corresponds to one of the choices to draw arrows on the Feynman diagram,
one of such choices is shown in Fig.~\ref{fig:se}(b).
The charge conservation imposes the following constraint:
\begin{eqnarray}
\sum\limits_{b = 2}^{2 N} \eta_b = 1 .
\label{charge}
\end{eqnarray}
This gives overall 
$C^N_{2 N - 1} = (2 N - 1)! / (N! (N - 1)!)$ 
terms in the sum over $\eta$. 
The long-distance asymptotics of the self-energy 
takes the form of Eq.~(\ref{Gred}):
\begin{eqnarray}
&& \Sigma_1(\tau, r) \to \left(\frac{k_1}{2 \pi r}\right)^{\frac{D - 1}{2}}  \nonumber \\
&&\times \left[i^{-\frac{D - 1}{2}}  e^{i k_1 r} \mathcal{S}_1(\tau, r) + i^{\frac{D - 1}{2}}  e^{-i k_1 r} \mathcal{S}_1(\tau, -r) \right], \label{Sigred}\\
&& \mathcal{S}_1(\tau, x) = \int\limits_{-\infty}^\infty \frac{d \delta p_1}{2 \pi} e^{i x \delta p_1} \Sigma_1(\tau, \delta p_1) , \label{sigonedim}
\end{eqnarray}
where $\mathcal{S}_1(\tau, x)$ is a one-dimensional self-energy.
In order to find $\mathcal{S}_1(\tau, x)$, we 
substitute the asymptotic expansion Eq.~(\ref{Gred}) of the Green functions 
in Eq.~(\ref{sigmageneral})
and compare with Eq.~(\ref{Sigred}):
\begin{eqnarray}
&& \!\!\!\!\!\!\!\! \mathcal{S}_1 (\tau, x) =  (-1)^{N - 1} c_1  \sum\limits_{\{\eta\}}  |\lambda_\eta|^2  \prod\limits_{a = 2}^{2 N}  \mathcal{G}_{a} (\eta_a \tau, s_a x) \mathcal{D} (x) ,
\label{sigmaxt} \\
&& \!\!\!\!\!\!\!\! \mathcal{D} (x) = \frac{e^{- i \mathcal{K} x}}{|x|^{2\nu}} 
i^{\sigma (D - 1) \sgn (x)}, \,\,\, \nu = \frac{1}{2}(N - 1) (D - 1) ,
\label{Dapp} 
\end{eqnarray}
where $s_a = -1$ ($ s_a = +1$) for $a \in \{2, \dots, N + \sigma\}$ ($a \in \{N + \sigma + 1, \dots , 2N \}$),
$|\mathcal{K}| \ll k_a$ and $\sigma$ are defined in Eq.~(\ref{reducedconserv})
and $c_1$ is the following constant:
\begin{eqnarray}
&& \!\!\! c_1 = \left(\frac{k_F}{k_1}\right)^{D - 1} \left(\frac{k_F}{2 \pi}\right)^{2 \nu} , \, 
\left(k_F\right)^{2 N} \equiv k_1 \cdot \dots \cdot k_{2 N} ,
\label{c1} 
\end{eqnarray}
where $k_F$ is the average Fermi momentum.
In Eq.~(\ref{sigmaxt}) we only retained slowly varying terms because we are interested 
in the long distance correlations.
Note that $1 / \mathcal{K}$ is the only 
length scale that survived the dimensional reduction.
The fluctuations at $|x| \gg 1 / |\mathcal{K}|$ are not important due to the oscillating exponential in Eq.~(\ref{Dapp}), 
and thus $1/|\mathcal{K}|$ defines the finite range of the interaction.
At $\mathcal{K} = 0$ there are no internal length scales that enter Eq.~(\ref{sigmaxt})
which results in the emergent long range order.

\section{Strong coupling limit}

In this section we consider
the strong coupling regime,
when  
the Green function close to the Fermi surfaces, see Eq.~(\ref{G}), 
is entirely defined by its self-energy:
\begin{eqnarray}
G_a (i \omega_a, \delta p_a) \approx - \frac{1}{\Sigma_a (i \omega_a, \delta p_a)} .
\label{green}
\end{eqnarray}
This approximation breaks down far away from the Fermi surfaces 
where the many-body scattering is off-resonant.
In other words, the LFL is restored in the 
ultraviolet limit far away from the Fermi surface.
Only the quantum states close to the Fermi surface
are strongly affected by the interaction.
These quantum states represent the infrared sector of the problem.

In the $(\tau, x)$ representation Eq.~(\ref{green}) takes the integral form:
\begin{eqnarray}
\!\!\!\!\!\!\!\! \delta (\tau) \delta (x) = - \int d\tau' dx' \, \mathcal{G}_a (\tau - \tau', x - x') \mathcal{S}_a (\tau', x') ,
\label{strongcoup}
\end{eqnarray}
where $\mathcal{G}_a(\tau, x)$ and $\mathcal{S}_a(\tau, x)$ are defined in Eqs.~(\ref{gonedim}) and (\ref{sigonedim}), respectively.
Substituting Eq.~(\ref{sigmaxt}) into Eq.~(\ref{strongcoup}) results in the integral Dyson equation for the Green function
calculated within the SCBA and in the strong coupling limit:
\begin{eqnarray}
&& \delta (\tau) \delta (x) = (-1)^N c_1 \int d\tau' dx' \, \mathcal{G}_1 (\tau - \tau', x - x')   \nonumber \\
&&\times \sum\limits_{\{\eta\}} |\lambda_\eta|^2 \prod\limits_{a = 2}^{2 N} \mathcal{G}_{a} (\eta_a \tau', s_a x') \mathcal{D} (x') .
\label{gequation}
\end{eqnarray}

At the FSR the effective interaction $\mathcal{D}(x)$ is quasi-long-range, i.e. it looks the same at all scales,
see Eq.~(\ref{Dapp}) at $\mathcal{K} = 0$.
In contrast to Eqs.~(\ref{G}) and (\ref{sigmageneral}) for the total Green function and the
self-energy, 
which contain the information about the physical scales
such as the Fermi momenta $k_a$ 
and corresponding band splittings,
Eq.~(\ref{gequation}) 
is free from any physical energy or length scale.
This observation suggests that the Green functions in Eq.~(\ref{gequation})
are also universal scaling functions.
As all $2 N$ Green functions occur in the product in Eq.~(\ref{gequation}), 
they are equivalent, and, thus,
we expect the same critical exponents for all of them regardless of the index $a$.
A simple dimensional analysis of Eq.~(\ref{gequation}) yields:
\begin{eqnarray}
\mathcal{G}_a (\tau, x) \propto \frac{1}{|\tau|^{2 h}} \frac{1}{|x|^{2 l}}, \,\,\,\,\, h = \frac{1}{2 N}, \, l = \frac{1 - \nu}{2 N} ,
\label{scalingapp}
\end{eqnarray}
where $\nu$ is given by Eq.~(\ref{Dapp}).
The time and coordinate scalings of the Green function 
 are different due to the effective interaction $\mathcal{D}(x)$.
This naturally suggests the separation of temporal and spatial dynamics:
\begin{eqnarray}
\mathcal{G}_a (\tau, x) = C_a g (\tau) \gamma (x) ,
\label{separation}
\end{eqnarray}
where $C_a$ is some constant that might be different for different $a$.
This ansatz is different from the truly one-dimensional case $D = 1$ in which the temporal and spatial scalings are the same and the separation argument does not apply for $D=1$.
Instead, left and right linear combinations of time and coordinate must be used for $D=1$.
Thus, our current analysis of Eq.~(\ref{gequation}) is only applicable for $D > 1$ where the effective interaction $\mathcal{D}(x)$ 
breaks the equivalence between time and coordinate.
The Dyson Eq.~(\ref{gequation}) then separates into a time and a coordinate equation:
\begin{eqnarray}
&& \!\!\!\!\!\!\!\!\!\! \delta (\tau) = (-1)^{N} \int d\tau' \, g (\tau - \tau') g(-\tau')^{N - 1} g(\tau')^N ,
\label{g}\\
&& \!\!\!\!\!\!\!\!\!\! \delta (x) \! = \! \!\! \int  \!\! dx' \gamma (x - x') \gamma(-x')^{N + \sigma - 1} \gamma(x')^{N - \sigma} \mathcal{D} (x') . \label{gamma}
\end{eqnarray}
As only the product of $g(\tau)$ and $\gamma(x)$ 
is important, see Eq.~(\ref{separation}),
we have an additional symmetry under the transformation $g(\tau) \to C g(\tau)$ and $\gamma(x) \to \gamma(x) / C$ for any complex $C\neq 0$.
The coefficients $C_a$ satisfy the following algebraic equation:
\begin{eqnarray}
c_1 \lambda^2 \prod\limits_{a = 1}^{2 N} C_a = 1 , \,\,\,\,\, \lambda = \sqrt{\sum\limits_{\{\eta\}}|\lambda_\eta|^2 } .
\label{Ca}
\end{eqnarray}

Equation~(\ref{Ca}) will be different 
if we consider the Dyson equation Eq.~(\ref{strongcoup}) for some other Fermi surface index $a \ne 1$.
This is due to the non-universal coefficient $c_1 \propto k_1^{1 - D}$, see Eq.~(\ref{c1}),
which is in general replaced by $c_a \propto k_a^{1 - D}$.
At the same time the product of all coefficients $C_a$ in Eq.~(\ref{Ca})
is independent of the Fermi surface index.
This 
apparent conflict
arises due to inadequacy of the long distance 
expansion, Eq.~(\ref{Gred}), at small 
distances $r \lesssim 1/k_F$,
$k_F$ is an average Fermi momentum.
Indeed, the right hand side of Eq.~(\ref{gequation}) has to diverge at $x = 0$
which corresponds to the ultraviolet divergence 
of the propagators $\mathcal{G}_a(\tau, x)$ at $x = 0$, see Eq.~(\ref{scalingapp}).
On the other hand, such a divergence at small $r \lesssim 1/k_F$  is regularized (and thus absent) in the exact Green function $G_a(\tau, r)$.
In other words, the long-distance infrared limit
that we take here does not allow us to define the exact 
numerical prefactor in Eq.~(\ref{gequation})
as it depends on the ultraviolet regularization.
At the same time,
the scale-independent limit of the Dyson Eq.~(\ref{gequation}) 
at $\mathcal{K} = 0$ implies that $c_1$ has to be replaced by a universal coefficient $c$ 
which we estimate via  dimensional analysis:
\begin{eqnarray}
&& c_1 \to c \sim \left(\frac{k_F}{2 \pi}\right)^{2 \nu} . \label{c}
\end{eqnarray}

We can choose all coefficients $C_a$ positive, i.e. $C_a > 0$.
Non-trivial phases can come from solutions for $g(\tau)$ and $\gamma(x)$ which we consider later.
All coefficients $C_a > 0$ are combined in the single product in Eq.~(\ref{Ca}) with $c_1$ replaced by $c$, see Eq.~(\ref{c}), so it is not possible to determine them separately without connecting the infrared interaction-dominated limit 
with the ultraviolet limit far away from the Fermi surfaces which is given by the LFL.
From Eq.~(\ref{Ca}) we conclude that the coefficients $C_a$ scale with the interaction strength as $C_a \propto \lambda^{-1/N}$.
This corresponds to the following self-energy scaling:
\begin{eqnarray}
&& \Sigma_a \propto \lambda^{1/N} . \label{siglam}
\end{eqnarray}
This scaling justifies the strongly interacting limit $\lambda \to \infty$ at which the self-energy correction is dominant over the single-particle spectral part.
Apart from the considered interaction scaling, the coefficients $C_a$ play no role in the infrared physics that we study here, 
so we can concentrate on the universal functions $g(\tau)$ and $\gamma(x)$.

We observe here that Eq.~(\ref{g}) is the Dyson equation for the generalized SYK model, 
referred to as $q$-SYK model~\cite{Ye,SachdevX}, with $q = 2 N$ in our case.
The SYK model describes $(0+1)$-dimensional strongly correlated fermions with all-to-all interactions whose matrix elements are randomly distributed.
Quite remarkably,  
the temporal dynamics for our $(D + 1)$-dimensional system without having any randomness in our model and for short-range interactions given by the Hamiltonian $V(R)$, see Eq.~(\ref{VN}),
maps exactly onto the $q$-SYK model with $q = 2 N$.
The SYK model is the central example 
of the AdS/CFT correspondence \cite{maldacena}
and plays an important role in understanding 
the nature of strange metals \cite{Parcollet,Chowdhury,Davison}.
In contrast to Eq.~(\ref{g}),
Eq.~(\ref{gamma})
contains the effective interaction 
$\mathcal{D}(x)$
which can be interpreted as a propagator of an emergent complex boson.
This emergent boson is merely a result of the dimensional reduction.
Its propagator $\mathcal{D}(x)$ 
contains the momentum mismatch $\mathcal{K}$ (which is zero at the FSR) and the geometric factors coming from the $s$-wave expansion of the fermion propagators, see Eqs.~(\ref{Gred}) and (\ref{Dapp}).

The Matsubara Green function $G_a (\tau, r)$, see Eq.~(\ref{Gred}), has to be 
real-valued which is evident from the spectral representation, see Eq.~(\ref{specrep}).
This is equivalent to the following constraint:
\begin{eqnarray}
&& \mathcal{G}_a^* (\tau, r) = \mathcal{G}_a (\tau, -r) ,
\end{eqnarray}
where $\mathcal{G}_a (\tau, r)$ is defined in Eq.~(\ref{gonedim}).
Together with the separation of variables, see Eq.~(\ref{separation}),
this results in the following conditions on the universal functions $g(\tau)$ and $\gamma(x)$:
\begin{eqnarray}
&& g^*(\tau) = g(\tau), \, \gamma^*(-x) = \gamma(x) . \label{sym}
\end{eqnarray}
In other words, 
$g(\tau)$ and $\gamma(\delta p)$ 
have to be real-valued functions.
Following Refs.~\cite{SachdevX,Parcollet},
we find the universal functions $g(\tau)$ and $\gamma(x)$:
\begin{eqnarray}
&& g(\tau) \propto \frac{\sgn(\tau)}{|\tau|^{2h}} , \label{gphsym} \\
&& \gamma(x) \propto \frac{1}{|x|^{2 l}} ,  \label{gamx}
\end{eqnarray}
where we used the symmetry of Eq.~(\ref{sym}).

Equation~(\ref{g})
can be easily extended to the case of finite temperature $T$ 
due to the emergent conformal symmetry, for details see Refs.~\cite{SachdevX,Parcollet}:
\begin{eqnarray}
&& g_T (\tau) \propto \sgn(\tau) \left|\frac{\pi T}{\sin \left(\pi T \tau\right)}\right|^{2 h} , \label{gT}
\end{eqnarray}
where $\tau \in (-1/T, 1/T)$ is the imaginary time, with the Boltzmann constant set to $k_B=1$.
Using Eqs.~(\ref{Gred}), (\ref{separation}), (\ref{gphsym}), and (\ref{gamx}),
we find explicitly the long time and long distance asymptotics 
of the finite temperature Matsubara Green function:
\begin{eqnarray}
&& \!\!\!\!\! G_a (\tau, r) \propto \sgn(\tau) \left|\frac{\pi T}{\sin \left(\pi T \tau\right)}\right|^{2 h} \frac{\cos \left(k_a r - \frac{\pi}{2}\frac{D - 1}{2}\right)}{r^{(D + 1) h}} , \label{greenasympt}
\end{eqnarray}
where $h = 1/(2 N)$ is the temporal scaling dimension.
The corresponding Fourier transform $G_a (i\omega_a, \delta p_a)$ becomes:
\begin{eqnarray}
&& \!\!\!\!\!\!\!\!\!\! G_a (i\omega_a, \delta p_a) \!=\! - \frac{i}{T}\! \left|\frac{T}{\Lambda}\right|^{2 h} \left|\frac{k_a}{\delta p_a}\right|^{1 - 2 l} \!\!\! \frac{\displaystyle \Gamma \left(h + \frac{\omega_a}{2 \pi T}\right)}{\displaystyle \Gamma \left(1 - h + \frac{\omega_a}{2 \pi T}\right)} , \label{matw}
\end{eqnarray}
where $\omega_a$ is the discrete Matsubara frequency, $k_a$  the Fermi momentum corresponding to the $a^{\rm th}$ Fermi surface.
Here we restored the physical dimensions
using Eqs.~(\ref{Lambda}), (\ref{siglam}).
The unknown numerical factors are hidden in the 
phenomenological ultraviolet scale $\Lambda$
that defines the interaction strength, see Eq.~(\ref{Lambda}).

Analytically continuing the Matsubara Green function 
Eq.~(\ref{matw}) to real frequencies,
we find the retarded Green function:
\begin{eqnarray}
&& \!\!\!\!\!\!\!\!\!\! G_a^{R} (\omega_a, \delta p_a) \! = \! - \frac{i}{T} \! \left|\frac{T}{\Lambda}\right|^{2 h} \! \left|\frac{k_a}{\delta p_a}\right|^{1 - 2 l} \!\!\! \frac{\displaystyle \Gamma \left(h - \frac{i\omega_a}{2 \pi T}\right)}{\displaystyle \Gamma \left(1 - h - \frac{i\omega_a}{2 \pi T}\right)} ,
\label{gret}
\end{eqnarray}
where $\omega_a$ is a real frequency.
The advanced Green function 
is complex conjugate to the retarded one.
In the limit of large frequency
$\omega \gg T$ 
we restore the power-law frequency scaling 
$G_a(\omega_a, \delta p_a) \propto |\omega_a|^{2 h - 1}$, where
$2 h - 1 < 0$ signaling the singularity at $\omega \to 0$.
At finite temperature $T$ 
this singularity is cut 
as $G_a (\omega_a = 0, \delta p_a) \propto T^{2 h - 1}$, see Eqs.~(\ref{matw}) and (\ref{gret}).

As we know the exact retarded Green function, 
we can calculate the fermion spectral density:
\begin{eqnarray}
&& \mathcal{A}_a (\omega_a, \delta p_a) = - \frac{1}{\pi} \mbox{\rm{Im}}\left[G_a^R(\omega_a, \delta p_a)\right] = \frac{\sin (\pi h)}{\pi^2 T}  \nonumber \\
&&\times \left|\frac{T}{\Lambda}\right|^{2 h} \left|\frac{k_a}{\delta p_a}\right|^{1 - 2 l} \cosh \left(\frac{\omega_a}{2 T}\right) \left|\Gamma\left(h + \frac{i \omega_a}{2 \pi T}\right)\right|^2 . \label{den}
\end{eqnarray}
The temperature and frequency dependence of
the spectral density correspond to 
the $q$-SYK, $q = 2N$, model, see Ref. \cite{Parcollet}.
The electron spectral function, see Eq.~(\ref{den}),
does not contain any sharply defined 
quasiparticle peaks.
Instead, there is the power law 
scaling in the momentum $\delta p_a$ and frequency $\omega_a \gg T$.
At low frequency $\omega_a \ll T$, the spectral density 
is a power-law function of temperature.
Thus, measuring the momentum, frequency, and temperature scaling of the spectral 
density close to the Fermi surface
allows one to identify experimentally the scaling dimensions $h$ and $l$ 
that determine the quantum critical point.

\section{Exactness of SCBA and stability of the quantum critical point}

In order to calculate the electron self-energy,
we used the SCBA, 
i.e. we neglected the interaction vertex corrections,
see Fig.~\ref{fig:se}.
Here we aim to show that the SCBA 
is exact in the strong coupling limit.
For this we investigate the symmetries of the Dyson Eqs.~(\ref{g}) and (\ref{gamma}).
It turns out that both equations are invariant under  time and coordinate reparametrizations which constitutes the 
two-dimensional conformal symmetry.
This means that if $g(\tau, \tau')$ [$\gamma(x, x')$]
are  solutions of the Dyson Eq.~(\ref{g}) [Eq.~(\ref{gamma})],
then the following functions $\tilde g (t_1, t_2)$ [$\tilde \gamma (\xi_1, \xi_2)$] 
are also solutions of the corresponding Dyson equations:
\begin{eqnarray}
&& \tau = f_1 (t), \, x = f_2(\xi) , \\
&& \tilde g (t_1, t_2) = \left|f_1'(t_1) f_1'(t_2) \right|^h g(f_1(t_1), f_1(t_2)) , \\
&& \tilde \gamma(\xi_1, \xi_2) = \left|f_2'(\xi_1) f_2'(\xi_2)\right|^{l} \gamma(f_2 (\xi_1), f_2(\xi_2)) ,
\end{eqnarray}
where $f_{1,2}(z)$ is a real-valued monotonic function, with
$f'_{1,2}(z)$ being its derivative with respect to $z$.
In contrast to the SYK Dyson Eq.~(\ref{g}),
the spatial Dyson Eq.~(\ref{gamma})
contains the function $\mathcal{D} (x)$ 
which transforms as a propagator of an emergent 
boson field with the scaling dimension $\nu$:
\begin{eqnarray}
&& \tilde \mathcal{D}(\xi_1, \xi_2) = \left|f_2'(\xi_1) f_2'(\xi_2)\right|^\nu \mathcal{D}(f_2(\xi_1), f_2(\xi_2)) .
\end{eqnarray}
The conformal symmetry of both temporal and spatial Dyson Eqs.~(\ref{g}) and (\ref{gamma})
is signaling an interaction fixed point in the renormalization group sense.
In other words, 
our initial ansatz that the interaction vertex 
is not renormalized in the infrared limit 
turned out to be correct in the strong coupling limit
due to the emergent conformal symmetry.

\begin{figure}[t]
	\includegraphics[width=\columnwidth]{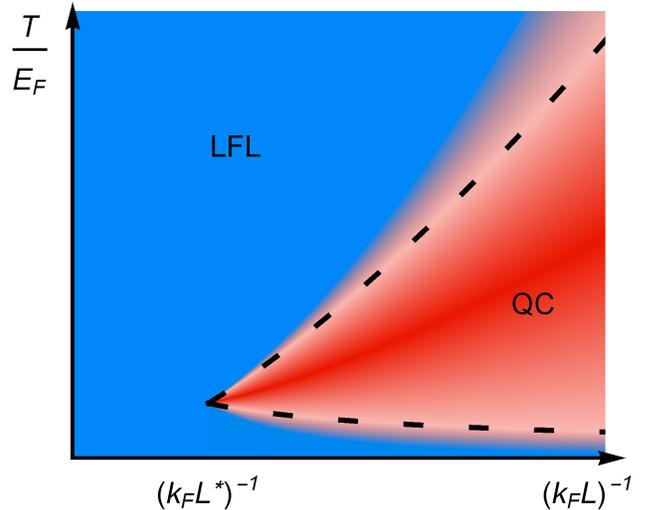}
	\caption{ 
		The $L$-$T$ phase diagram illustrating the crossover 
		between the LFL (blue) and quantum critical (QC) state (red).
		The dashed lines correspond to upper and
		lower bounds in Eq.~(\ref{Trange}).
		The upper bound for the wire length 
		$L^*$ corresponds to Eq.~(\ref{Lstar}).
	}
	\label{fig:LT}
\end{figure}

So far, we have solved exactly
the problem of a strongly interacting electron gas that is
tuned to the FSR. 
However, we neglected the single-particle spectral part 
compared to the self-energy, see Eq.~(\ref{green}).
The strongly interacting limit is justified
if $|\Sigma_a (i \omega_a, \delta p_a)| \gg \mbox{\rm{max}} \{\omega_a, v_a \delta p_a\}$, see Eq.~(\ref{G}),
where $v_a$ is the Fermi velocity 
at the $a^{\rm th}$ Fermi surface.
Here we consider the case of finite temperature $T$ and finite size $L$ of the system.
In this case the infrared and long distance limits correspond to $\omega_a \sim T$ and $\delta p_a \sim 1/L$.
Using Eqs.~(\ref{green}) and (\ref{matw}),
we can estimate the self-energy at the smallest frequencies and momenta:
\begin{eqnarray}
&& \Sigma_a (i \omega_a \sim i T, \delta p_a \sim 1/L) \propto \Lambda^{2 h} T^{1 - 2 h} (k_a L)^{2 l - 1} .
\end{eqnarray}
Therefore, 
the strong coupling limit corresponds to 
the following constraint:
\begin{eqnarray}
&& \Lambda^{2 h} T^{1 - 2 h} (k_F L)^{2 l - 1} \gg \mbox{\rm{max}} \left\{T, \frac{v_F}{L} \right\} , \label{cond}
\end{eqnarray}
where we introduced the average Fermi momentum $k_F$
and average Fermi velocity $v_F$.
This condition can be represented in 
simpler form:
\begin{eqnarray}
&& \frac{r_s^{N - 1}}{(k_F L)^{\frac{1 - 2 l}{2 h}}} \gg \frac{T}{E_F} \gg  \frac{1}{r_s} \left(k_F L\right)^{- \frac{2 l}{1 - 2 h}}, \label{Trange}
\end{eqnarray}
where we used Eq.~(\ref{Lambda}) 
to express everything in terms of $r_s \gg 1$.
Here it is important that 
$1 - 2 l > 0$ for any $N \ge 2$, $D > 1$,
see Eq.~(\ref{scalingapp}).
This results in an upper bound $L^*$ for the 
system size $L$:
\begin{eqnarray}
&& L \ll L^* = \frac{1}{k_F} r_s^{\frac{1 - 2 h}{1 - 2h - 2 l}} . \label{Lstar}
\end{eqnarray}
Thus,
the quantum critical point corresponding to $\mathcal{K} = 0$
is stable 
only for finite size $L \ll L^*$ 
and in the temperature range given by Eq.~(\ref{Trange}),
see Fig.~\ref{fig:LT}.

In the dimensional reduction we rely on the rotational symmetry of the Fermi surfaces.
However, main conclusions of our paper will still be true in case of small asymmetry
of the Fermi surfaces.
Let the Fermi momentum of the $a^{\rm th}$ Fermi surface range from $k_a - \Delta k_a$ to $k_a + \Delta k_a$, where 
$\Delta k_a \ll k_a$.
Then in addition to Eq.~(\ref{Trange})
the temperature must also satisfy
$T \gg v_a \Delta k_a$. Here, $v_a$ is the Fermi velocity of the $a^{\rm th}$ Fermi surface,
so that the Fermi surface asymmetry is smoothened out by thermal effects.
Comparing with Eq.~(\ref{Trange}) this gives 
an upper bound for $\Delta k_a$:
\begin{eqnarray}
&& \frac{v_a \Delta k_a}{E_F} \ll \frac{r_s^{N - 1}}{(k_F L)^{\frac{1 - 2 l}{2 h}}} .
\end{eqnarray}

\section{Linear response functions}

In this section 
we consider the linear response functions 
that can also be used for experimental
identification \cite{Stano}
of the predicted quantum critical state.
In case of non-degenerate Fermi surfaces 
the linear response functions are represented by the 
following charge susceptibilities
that we calculate within the Matsubara technique:
\begin{eqnarray}
\chi_{ab} (\tau, r) = - G_a (\tau, r) G_b (-\tau, r),
\label{chi}
\end{eqnarray}
where $a$ and $b$ are the Fermi surface indices,
$r$ is an absolute value of a $D$-dimensional 
coordinate vector,
$\tau$ is imaginary time.
The vertex correction 
is neglected in Eq.~(\ref{chi}).
In Appendix A we show that the vertex correction
to a linear response function 
does not affect the scaling properties.
Using the long time and long distance asymptotics of the finite temperature Matsubara Green function
which is given by Eq.~(\ref{greenasympt}),
we find the asymptotics of the linear response functions:\\
\begin{eqnarray}
&& \chi_{ab} (\tau, r) \propto \left|\frac{\pi T}{\sin (\pi T \tau)}\right|^{4 h}  \nonumber \\
&&\hspace{40pt} \times \frac{\cos\left(k_+ r - \pi \frac{D - 1}{2}\right) + \cos\left(k_- r\right)}{r^{2 (D + 1) h }} , \label{chias}
\end{eqnarray}
where $k_\pm = k_a \pm k_b$.
The temporal part coincides with the SYK susceptibility,
see Ref.~\cite{Parcollet}.
The spatial part is dominated by the Kohn anomalies \cite{Kohn}.
In case $a = b$ the Matsubara susceptibility $\chi_{aa}(\tau, r)$ has a non-oscillatory quasi-long-range component.
The Fourier transform of Eq.~(\ref{chias}) yields $\chi_{ab} (i\omega, q)$:
\begin{eqnarray}
\chi_{ab} (i\omega, q) \propto \chi_M (i\omega) \chi_{ab} (q) , \label{chiqw}
\end{eqnarray}
where
\begin{eqnarray}
 \chi_M (i\omega) = \frac{\Gamma(1 - 4 h)}{\Lambda} \left|\frac{T}{\Lambda}\right|^{4 h - 1} \frac{\displaystyle \Gamma\left(2 h + \frac{ \omega}{2 \pi T}\right)}{\displaystyle \Gamma\left(1 - 2 h + \frac{\omega}{2 \pi T}\right)}  \label{chim}
\end{eqnarray}
and
\begin{eqnarray}
&& \chi_{ab} (q \approx |k_\pm|) = \frac{(k_a k_b)^{D - (D + 1) h} \Gamma\left(\frac{D + 1}{2} (1 - 4 h)\right)}{\displaystyle q^{\frac{D - 1}{2}} |q - |k_\pm||^{\frac{D + 1}{2} (1 - 4 h)}}  \nonumber \\
&& \times \cos\left(\frac{\pi}{4} \left[(D + 1) (1 - 4 h) \pm \sgn (q - |k_\pm|) (D - 1)\right]\right).\,\,\,\,\,\,\,\,\,\, \label{chiq}
\end{eqnarray}
Here, the index $M$ in $\chi_M (i\omega)$ stands for Matsubara susceptibility,
and $\omega$ is the bosonic Matsubara frequency.
We also restored the physical dimensions
using Eq.~(\ref{matw}),
even though the overall numerical factor is suppressed.
Equation~(\ref{chiq}) is also valid in the case 
$a = b$ with $k_- = 0$ where $\chi_{aa}(i\omega, q)$ is divergent for $q \to 0$.
Analytically continuing $\chi_M(i\omega)$ 
to real frequencies we find the retarded linear response function:
\begin{eqnarray}
&& \chi_R (\omega) = \frac{\Gamma(1 - 4 h)}{\Lambda} \left|\frac{T}{\Lambda}\right|^{4 h - 1} \frac{\displaystyle \Gamma\left(2 h - \frac{i \omega}{2 \pi T}\right)}{\displaystyle \Gamma\left(1 - 2 h - \frac{i \omega}{2 \pi T}\right)} ,  \,\,\,\,\, \label{chiret} 
\end{eqnarray}
where $\omega$ is a real frequency.
The imaginary part of the retarded linear response, 
describing the dissipation due to interaction effects, becomes:
\begin{eqnarray}
&& {\rm Im} \left(\chi_{ab}^R (\omega, q)\right) \propto \chi_{ab}(q) {\rm Im} \left(\chi_R (\omega)\right) , \\
&& {\rm Im} \left(\chi_R (\omega)\right) = \cos(2 \pi h) \frac{\Gamma(1 - 4 h)}{\pi \Lambda} \left|\frac{T}{\Lambda}\right|^{4 h - 1}  \nonumber\\
&& \times \sinh \left(\frac{\omega}{2 T}\right) \left|\Gamma\left(2 h + \frac{i \omega}{2 \pi T}\right)\right|^2 , \label{imchi}
\end{eqnarray}
where $\chi_{ab} (q)$ is given by Eq.~(\ref{chiq}).

For $N = 2$ we have $h = 1/4$,
which results in a divergence in Eq.~(\ref{chiqw}).
Regularizing this divergence,
we find the retarded susceptibility 
for $h = 1/4$:
\begin{eqnarray}
&& \!\!\!\!\! \chi_R (\omega) = \frac{1}{\Lambda} \left[\ln \left(\frac{\Lambda}{2 \pi T}\right) - \psi\left(\frac{1}{2} - \frac{i \omega}{2 \pi T}\right)\right], \, h = \frac{1}{4} , \label{quarter} \\
&& \!\!\!\!\! {\rm Im} \left(\chi_R (\omega)\right) = \frac{\pi}{2 \Lambda} \tanh\left(\frac{\omega}{2 T}\right) .
\end{eqnarray}
Here $\psi(z) = \Gamma'(z)/\Gamma(z)$ is the digamma function.
The real part of $\chi_R (\omega)$ is 
logarithmically divergent at $\omega \gg T$.
These equations correspond to the standard SYK susceptibility, see Ref.~\cite{Parcollet}.
For $N = 2$ the Kohn anomalies are 
logarithmically divergent at $q \approx |k_\pm| = |k_a \pm k_b|$:
\begin{eqnarray}
&& \!\!\!\!\! \!\!\!\chi_{ab} (q \approx |k_\pm|) \propto (k_a k_b)^{\frac{D}{2}} \left|\frac{k_a k_b}{q^2}\right|^{\frac{D - 1}{4}} \ln\left|\frac{k_a k_b}{(q -|k_\pm|)^2}\right| . \label{chiq2}
\end{eqnarray}
From Eq.~(\ref{chiq2}) we see that 
the susceptibility at $a = b$ and $q \to 0$ 
diverges as a power law 
with additional logarithmic divergence $\chi_{aa} (q \to 0) \to \ln(k_a/q) / q^{(D - 1)/2}$.

The divergent Kohn anomalies given by Eq.~(\ref{chiq})
and the one-dimensional character of the radially nested scattering resonance may lead to spatially inhomogeneous orders.
For example, coupling to intrinsic phonons 
can result in a charge density order due to the Peierls instability \cite{Peierls}.
Another example comes from the Ruderman-Kittel-Kasuya-Yoshida (RKKY) exchange interaction between  magnetic impurities which is mediated by itinerant fermions \cite{Ruderman}.
Magnetic helical order can be established if the spin susceptibility of the itinerant fermions has a divergent Kohn anomaly \cite{Braunecker,Scheller}.

Next, we  derive the optical conductivity
which takes the following form
in the Matsubara representation \cite{Parcollet,Georges}:
\begin{eqnarray}
&& \sigma_{\alpha \beta} (i \omega, q) \nonumber \\
&& = \frac{1}{\omega} \sum\limits_{a, \bm p} T \sum\limits_{\varepsilon} v_a^\alpha v_a^\beta G_a (i \varepsilon, \bm p) G_a (i (\varepsilon + \omega), \bm p + \bm q) , \label{sig}
\end{eqnarray}
where we sum over the fermionic Matsubara frequencies $\varepsilon$
and $v_a^\alpha$ is the $\alpha$ component of the 
current vertex.
In Appendix A we argue that 
the current vertex renormalization 
does not affect the scaling in the conformal limit, i.e. we 
can use Eq.~(\ref{sig}) 
with current vertices represented by the corresponding Fermi velocity $\bm v_a$.
Due to the separation of variables in the Green function,
the $\omega$-dependent part of the Matsubara optical conductivity 
is simply given by $\chi_M (i\omega) / \omega$,
for $\chi_M(i \omega)$ see Eq.~(\ref{chim}),
while the $q$-dependent part is given by a power law
because $G_a (i \omega, \delta p_a) \propto 1 / |\delta p_a|^{1 - 2l}$, see Eq.~(\ref{matw}).
Analytically continuing to  real frequencies,
we find the retarded optical conductivity:
\begin{eqnarray}
&& \sigma_{\alpha \beta} (\omega, q) = \delta_{\alpha \beta} \sigma(\omega, q) , \,\,\,\, \sigma(\omega, q) \propto  \frac{\chi_R(\omega)}{i \omega}  \frac{1}{q^{1 - 4 l}} , \label{conduct}
\end{eqnarray}
where $\delta_{\alpha\beta}$ appeared 
after the angular average over $v_a^\alpha v_a^\beta$,
$\omega$ is a real frequency,
and $l$ is the spatial scaling dimension, see Eq.~(\ref{scalingapp}).
We see from Eq.~(\ref{conduct}) that the dissipative real part of the optical
conductivity is proportional 
to the imaginary part of the retarded 
susceptibility:
\begin{eqnarray}
&& {\rm Re} (\sigma (\omega, q)) \propto \frac{{\rm Im} \left(\chi_R(\omega)\right)}{\omega} \frac{1}{q^{1 - 4 l}} . 
\end{eqnarray}

In the limit $q \to 0$ the optical conductivity diverges due to the 
translational invariance in our system.
The translational symmetry can be broken by 
either spatially varying electric field that results in a finite $q$ 
or by the finite size of the system (with open boundary conditions) which 
results in the infrared cut-off at $q \sim 1/L$,
$L$ is the system size.
In this work we do not consider the effects of
momentum non-conserving interactions like
disorder or Umklapp
scattering, i.e. we assume $L \ll \ell$,
where $\ell$ is the mean free path.
In such a clean limit $L \ll \ell$ the dc resistivity $\rho_{dc}(T, L)$ exhibits anomalous scaling with temperature and system size:
\begin{eqnarray}
&& \rho_{dc}(T, L) = \frac{1}{{\rm Re} (\sigma (\omega = 0, q = 1/L))} \propto \frac{T^{2 - 4 h}}{ L^{1 - 4 l}}. \label{rhodc}
\end{eqnarray}
As $1 - 4 l > 0$, the resistivity $\rho_{dc}(T, L)$ tends to zero for $L \to \infty$,
as it should in absence of  momentum non-conserving interactions.
Thus, measuring the temperature and sample size dependence for $L \ll \ell$ of the static resistivity
gives both the temporal $h$ and spatial $l$ 
conformal dimensions.

The case of $N = 2$ is particularly interesting.
In this case, the dc resistivity is linear in temperature $\rho_{dc} \propto T$,
see Eq.~(\ref{rhodc}). 
This is the characteristic feature of strange metals which has been observed experimentally in cuprates \cite{Daou,Keimer} and heavy fermion metals \cite{Gegenwart}.
Current theories of strange metals \cite{Parcollet,Davison,Chowdhury} are based on the SYK model \cite{Ye} that requires long-range interaction and random distribution of the interaction matrix elements.
In our model the effective interaction is short range, see Eq.~(\ref{VN}).
Moreover, there is no randomness involved in the problem.
Therefore, the new physical mechanism of the quantum criticality based on the resonant many-body exchange scattering that we propose in our study might play an important role in understanding the nature of strange metals.\\

\section{Conclusions}

In our study we theoretically discovered a novel physical state of strongly interacting fermions which can be realized in materials with multiple Fermi surfaces that are subject to a special resonant condition, FSR, given by Eq.~(\ref{reducedconserv}).
This phase can be experimentally identified by the spectral function showing no Landau quasiparticles close to the Fermi surface, by the anomalous power-law temperature and size dependence of the dc resistivity, 
and by the divergent susceptibilities in the static limit and at the Kohn anomalies.
We believe that the new exotic phase that we predict in this work can be found, for instance, in semiconductor heterostructures because of the large interaction parameter $r_s \gg 1$.
Moreover, the high quality and tunability of  semiconductor devices makes it possible to tune the system to the FSR, see Eq.~(\ref{reducedconserv}), which is required for establishing  the new quantum critical state. \\

\section{ACKNOWLEDGEMENTS}
We thank Oleg P. Sushkov
for inspiring and stimulating discussions.
We acknowledge support by the Georg H. Endress foundation, 
the Swiss National Science Foundation, and NCCR QSIT. This project received funding from the European Union's Horizon 2020 research
and innovation program (ERC Starting Grant, grant
agreement No 757725).

\appendix

\section{LINEAR RESPONSE FUNCTIONS}

In this section we make use of the conformal symmetry to demonstrate that the 
vertex corrections of the linear response functions, such as the dc conductivity and the charge susceptibilities, do not influence the critical exponents.
We consider the general case of linear response function $\chi_{AB} (t, t')$, where $A$ and $B$ are some operators that are placed in the vertices of the response function.
The dependence of $\chi_{AB}$ on the effective spatial coordinates $x, x'$ can be considered analogously.
Here we use the relation between the linear response functions and the four-point Green function $G^{(4)} (t_1, t_2; t_3, t_4)$:
\begin{eqnarray}
&& \chi_{AB} (t, t') = \mbox{\rm{Tr}} \left(A \, G^{(4)}(t, t; t', t') B\right) ,
\label{chiab}
\end{eqnarray}
where $\mbox{\rm{Tr}}$ stands for the trace over the index space.
It is not important for us how exactly the trace is taken,
here we are after the time scaling of $\chi_{AB}(t, t')$.
The global conformal symmetry (it consists of the translations, dilatations, and special conformal transformations) restricts the four-point Green function to the following form \cite{Ginsparg}:
\begin{eqnarray}
&& G^{(4)} (t_1, t_2; t_3, t_4) = F(\tau) \prod\limits_{i < j} t_{ij}^{- \frac{2 h}{3}} , \,\,\, \tau = \frac{t_{13} t_{24}}{t_{14} t_{23}} ,
\label{fourpoint}
\end{eqnarray}
where $t_{i j} = t_i - t_j$, $\tau$ is the conformal cross-ratio,
$F(\tau)$ is some function of the cross-ratio,
$h$ is the temporal conformal dimension, see Eq.~(\ref{scalingapp}).
Equation~(\ref{fourpoint}) explicitly separates the pairwise singularities when $t_i \to t_j$.
The scaling of $t_{ij}$ 
can be found from applying the rescaling of times $t_i \to s t_i$. 
On the one hand, each of the four fields in $G^{(4)}$ has conformal dimension $h$, so $G^{(4)}$ acquires the factor $s^{-4 h}$.
On the other hand, the factor $s$ comes from each of the six $t_{ij}$ which fixes their power to $-2h/3$.
In order to calculate $\chi_{AB} (t, t')$, we have to put $t_1 = t_2 = t$, $t_3 = t_4 = t'$, see Eq.~(\ref{chiab}).
This leads to  singularities in Eq.~(\ref{fourpoint}) since $t_{12} = t_{34} = 0$.
This problem can be avoided by setting small non-zero $t_{12}$ and $t_{34}$ and express them in terms of non-zero $t_{13} = t_{14} = t_{23} = t_{24} = t - t'$:
\begin{eqnarray}
&& t_{12} t_{34} = \frac{\tau - 1}{\tau} t_{13} t_{24} = \frac{\tau - 1}{\tau} (t - t')^2 .
\end{eqnarray}
Then we substitute it in Eq.~(\ref{fourpoint}) and take the limit $t_{12} \to 0$, $t_{34} \to 0$ which is after all equivalent to the limit $\tau \to 1$:
\begin{eqnarray}
&& \!\!\!\!\!\!\!\! G^{(4)} (t, t; t', t') \! = \! \lim\limits_{\tau \to 1} \!\! \left(\!\! F(\tau) \!\! \left(\! \frac{\tau - 1}{\tau} \right)^{-\frac{2 h}{3}} \!\right) \!\! (t - t')^{- 4 h} .
\label{G4scale}
\end{eqnarray}
The limit at $\tau = 1$ yields some constant that we are not interested in.
An important consequence of Eq.~(\ref{G4scale}) is the universal scaling of the linear response function with time, $\chi_{AB}(t, t') \propto |t - t'|^{-4h}$.
The dependence of the linear response function on the effective radial coordinate can be deduced similarly.
So, the conformal symmetry allows us to restore the time and coordinate scaling of any linear response function at zero temperature $T = 0$:
\begin{eqnarray}
 &&  \chi_{AB}(t, x, t', x') \propto |x - x'|^{- 4 l} |t - t'|^{- 4 h} ,
\label{chisc}
\end{eqnarray} 
where $l$ and $h$ are the spatial and the temporal conformal dimensions, respectively, see Eq.~(\ref{scalingapp}).
Note that this scaling is entirely determined by the conformal symmetry.
In particular, $\chi_{AB}(t, x, t', x')$ contains all corrections to the linear response vertex.

It is worth mentioning that the linear response function with all  vertex corrections neglected has exactly the same scaling:
\begin{eqnarray}
&& \chi^{(0)}_{AB} (t, x, t', x') \nonumber \\
&& = \mbox{\rm{Tr}} \left(G(t - t', x - x') A \, G(t'- t, x' - x) B\right) \nonumber \\
&& \propto |x - x'|^{- 4 l} |t - t'|^{-4 h} .
\label{chi0}
\end{eqnarray}
In this case the scaling is given directly by the two-point Green function, see Eqs.~(\ref{separation}), (\ref{gphsym}), and (\ref{gamx}).
This allows us to conclude that the corrections to the linear response vertices do not influence the time and coordinate scaling of the linear response function.
The vertex correction is also not important at 
finite temperature $T$
because it corresponds to a certain conformal transformation,
see Ref.~\cite{SachdevX}.

\section{COMMENTS ON THE ONE-DIMENSIONAL CASE}
All the results that we provide in the main text are only valid for $D > 1$ due to the separation of temporal and spatial dynamics, see Eq.~(\ref{separation}).
In case of $D = 1$ one has to figure out appropriate linear combinations of $x$ and $\tau$.
Usually, these combinations correspond to the left and right movers in real time 
and to complex coordinates $x \pm i \tau$ in the imaginary time \cite{Ginsparg}.

In the truly one-dimensional case one has to bosonize the effective interaction $V(R)$ given by Eq.~(\ref{VN}).
We recall that $V(R)$ contains $C_{2 N}^N$ different terms half of which are conjugates.
Right at the FSR that is given by Eq.~(\ref{reducedconserv}) $V(R)$ contains slowly varying terms that can be combined into $C_{2 N}^N$ non-commuting cosines.
At this point it is not exactly clear how to proceed with such a large number of non-commuting interaction terms.
However, due to the competing nature of these cosines we also expect a highly non-trivial quantum critical phase in this case.
The case of two non-commuting cosines corresponds to the self-dual sine Gordon model that describes parafermions \cite{Leche}.
Exactly solvable extensions of the sine Gordon model are typically built via extending the underlying symmetry group \cite{Teo,Kogan}.

\end{document}